\documentclass[12pt,preprint]{aastex}

\begin{document}
\title{Explosive Nucleosynthesis from GRB and Hypernova Progenitors:  Direct Collapse versus Fallback}

\author{Christopher L. Fryer\altaffilmark{1,2}, Patrick
A. Young\altaffilmark{2,3}, Aimee L. Hungerford\altaffilmark{4}}

\altaffiltext{1}{Department of Physics, The University of Arizona,
Tucson, AZ 85721} 
\altaffiltext{2}{Theoretical Division, LANL, Los Alamos, NM 87545}
\altaffiltext{3}{Steward Observatory, The University of Arizona,
Tucson, AZ 85721} 
\altaffiltext{4}{Computational and Computer Science Division, 
LANL, Los Alamos, NM 87545}

\begin{abstract}

The collapsar engine behind long-duration gamma-ray bursts extracts
the energy released from the rapid accretion of a collapsing star onto
a stellar-massed black hole.  In a collapsing star, this black hole can 
form in two ways:  the direct collapse of the stellar core into a black 
hole and the delayed collapse of a black hole caused by fallback in a 
weak supernova explosion.  In the case of a delayed-collapse black hole, 
the strong collapsar-driven explosion overtakes the weak supernova 
explosion before shock breakout, and it is very difficult to distinguish 
this black hole formation scenario from the direct collapse scenario.  
However, the delayed-collapse mechanism, with its double explosion, 
produces explosive nucleosynthetic yields that are very different 
from the direct collapse scenario.  We present 1-dimensional studies 
of the nucleosynthetic yields from both black hole formation scenarios, 
deriving differences and trends in their nucleosynthetic yields.

\end{abstract}


\section{Introduction}

Since the observation of SN 1997ef \citep{Iwa00}, the number of strong
supernovae has grown considerably.  This class of supernova, termed
hypernova, includes the supernovae associated with gamma-ray bursts
(see Maeda \& Nomoto 2003; Nomoto et al. 2004 for reviews).  Indeed,
one of the best observed examples of a hypernova is SN 1998bw, which
occurred concurrently and cospatially with Gamma-Ray Burst 980425
\citep{Gal98}.  It is believed that long-duration gamma-ray bursts
(GRBs) make up a subset (perhaps even the entire set) of all
hypernovae and it is often assumed that the engine driving all of
these bursts is the same ``collapsar'' engine \citep{Pods04}.

The collapsar engine is powered by the gravitational energy released
by matter accretion onto a stellar mass black hole, presumably from
the collapse of a massive star (Woosley 1993).  To extract this
energy, this matter must hang up in a disk prior to moving into the
event horizon of the black hole.  Two mechanisms have been proposed to
extract this accretion disk energy: neutrino annihilation and magnetic
field eruptions (Narayan et al. 1992). This energy drives a strongly
asymmetric explosion that develops into a strongly relativistic jet in
the case of long-duration GRBs.  Although the explosion 
is not strongly relativistic outside of the jet axis, it is likely 
that the entire star is disrupted by the explosion.

At the crux of this mechanism is the fact that the star collapses to a
black hole.  Stars can form black holes through two separate paths:
the direct collapse of a star down to a black hole or the fallback of
considerable material onto a proto-neutron star formed in a weak
supernova explosion (Fryer 1999). Fryer \& Kalogera (2001) argued that
this latter scenario could well be the most common path behind black
hole formation.  Since the mechanisms (neutrino annihilation and
magnetic fields) that extract the gravitational potential energy
released in the disk do not require that the compact remnant be a
black hole, it might be possible to produce hypernovae by accretion
onto neutron stars if the rate of accretion is sufficiently high
($\gtrsim 0.1 M_\odot s^{-1}$).  In this paper, we will consider not
only the direct and fallback collapse black-hole engines, but also
these rapidly accreting neutron star scenarios.

Most of the current models for nucleosynthetic yields of hypernovae
effectively assume that the black hole is formed in the direct
collapse of a black hole (e.g. Nomoto et al. 2005a,2005b and
references therein).  But with the possibility that a majority of
hypernovae could be produced by fallback black-holes, we have
a new paradigm for estimating the nucleosynthetic yields of these
explosions.  We will show in this paper that the nucleosynthetic yield
from a single strong explosion is quite different from that of a weak
supernova explosion followed by fallback and then a strong explosion.
In \S 2, we show the results of our hypernova explosions and review
the differences in the temperature and density evolution of ejecta for
single strong explosions versus weak explosions followed after a range
of delays by a strong explosion.  In \S 3, we present the yields of
these explosions.  We conclude with a discussion of the primary trends
in the yields, outlining the implications the yields have on hypernova
progenitors and ultimately stellar collapse.

\section{Hypernova Explosions}

\subsection{Numerical Setup and Single and Double Explosions}

We focus our study on 2 potential hypernova progenitors: a
23\,M$_\odot$ star that loses its hydrogen envelope in a binary
interaction prior to helium ignition (Case B mass transfer) and a
40\,M$_\odot$ single star (Young et al. 2006). Both observations
\citep{hjo03,sta03} and theoretical considerations \citep{woo93,zwm03}
call for a progenitor with no hydrogen envelope. These progenitors
represent the two primary formation channels for a hydrogen poor single star
collapsar model (as opposed to a merger-induced GRB). We follow the
evolution from collapse of the entire star through the bounce and
stall of the shock using the 1-dimensional core-collapse code
described in Herant et al.(1994) and Fryer et al. (1999).  This code
models the transport of 3 neutrino species
($\nu_e,\bar{\nu}_e,\nu_{\mu,\tau}$) using single energy flux-limited
diffusion and takes advantage of a 4-part equation of state that
covers a range of densities from above nuclear to ideal gas regimes
and includes both a small nuclear network and nuclear statistical
equilibrium prescriptions (see Herant et al. 1994 and Fryer et
al. 1999 for details).  At the end of this phase, the proto-neutron
stars of the 23 and 40\,M$_\odot$ stars have reached masses of 1.37
and 1.85\,M$_\odot$, respectively.

At this point in the simulation, we remove the neutron star core and
drive an explosion by artificially heating the 15 zones ($\sim
3\times10^{-2}$\,M$_\odot$) just above the proto-neutron star.  The
amount and duration of the heating is altered to achieve the desired
explosion energy, where explosion energy is defined as the total final
kinetic energy of the ejecta (see Table 1 for details).  The neutron
star is modeled as a hard surface.  Figure 1 shows the velocity
profiles of our strong explosions ($13,14 \times 10^{51}$\,erg for the
23,40\,M$_\odot$ respectively).  Note that the initial velocities are
above 30,000\,km\,s$^{-1}$ and even as the shock is breaking out of
the star, the peak velocity is above 20,000\,km\,s$^{-1}$ for both
models.  There is no fallback for these models.  This explosion mimics
the conditions for those gamma-ray bursts that are produced by the
direct collapse of the star to a black hole.

The velocity profile of our weak supernova explosions ($0.25,0.14
\times 10^{51}$\,erg for the 23,40\,M$_\odot$ respectively) is shown
in Figure 2.  Although the shock rises to peak velocities of
10,000\,km\,s$^{-1}$, by the time the shock breaks out of the star,
the peak velocities are less than 5,000-7,000\,km\,s$^{-1}$.
Notice also in Figure 1 that the inner material slows down as its
energy is used to accelerate the material above it.  This material
decelerates below the escape velocity and falls back onto the neutron
star.  This is the primary mechanism for supernova fallback (see Fryer
et al. 2006 for a review).  As this material falls back onto the
neutron star its density increases.  When the density of a fallback
zone rises above $5 \times 10^8\,{\rm g \, cm^{-3}}$, we assume that
neutrino cooling will quickly cause this material to accrete onto the
neutron star and we remove the zone from the calculation (adding its
mass to the neutron star).  The mass of the neutron star as a function
of time is shown in Fig. 3.  Be aware that even if no gamma-ray burst
is produced, much of this material need not accrete onto the neutron star.
Especially if this material has angular momentum, it is likely that
some of it will be ejected in a second explosion. (In a case with no
GRB this will presumably be due to a similar mechanism, but without
the very high Lorentz factors.) This ejecta is a site of heavy element
production (Fryer et al. 2006).  As we focus here only on the
explosive nucleosynthesis, we will not discuss this ejecta further.

The fallback black hole scenario corresponds to an initially weak
explosion as is seen in Figure 2 that is then followed by a strong
explosion.  Without knowing the angular momenta of these stars
(accurate angular momentum estimates are beyond the capability of
current stellar evolution codes), it is difficult to determine when a
disk will form that powers the collapsar engine.  Hence, we consider a
range of delay times corresponding to the onset of mass fallback onto
the neutron star and to the timescale when the remnant mass exceeds 2
different limits (these limits are progenitor specific - see table 1).
We produce the second explosion by taking the data from the weak
explosion at a given time delay and then re-injecting energy into the
base of the star (again, the inner $\sim 3\times10^{-2}$\,M$_\odot$)
and driving an explosion with energy in excess of $10^{52}$\,erg.

\subsection{Second Temperature and Density Histories}

The final nucleosynthetic yield of the ejected matter is set by the
initial abundance (in particular the neutron excess), but even more
importantly by the density and temperature evolution of the ejecta.
We must understand this evolution before we can understand the final
yields of these explosions.  Figure 4 shows the temperature evolution
of 2 mass zones in our explosion models.  Figure 5 shows the density
evolution for these same zones.  Note that the structure changes
dramatically for the different explosion models.

The behavior in these figures can easily be understood by using shock
jump conditions to determine the density ($\rho_{\rm shock}$) and
pressure ($P_{\rm shock}$) after the shock has passed (e.g. Chevalier
1989):
\begin{equation}
\rho_{\rm shock} = \frac{(\gamma + 1) M^2}{(\gamma - 1) M^2 + 2}\rho_0
\end{equation}
\begin{equation}
P_{\rm shock} = (2 \rho_0 v_{\rm shock}^2 - (\gamma - 1) P_0)/(\gamma + 1)
\end{equation}
where $\gamma$ is the adiabatic, $M$ and $v_{\rm shock}$ are the Mach
number and velocity respectively of the shock, and $\rho_0, P_0$ are
the initial density and pressure profiles.  In the strong shock limit,
these equations reduce to familiar limits:
\begin{equation}
\rho_{\rm shock} = (\gamma + 1)/(\gamma - 1) \, \rho_0
\end{equation}
\begin{equation}
P_{\rm shock} = (2 \rho_0 v_{\rm shock}^2)/(\gamma + 1).
\end{equation}
In the strong shock limit we expect the pressure, and hence the
temperature, to depend upon the velocity of the shock, but the density
is increased by a constant factor (independent of the shock velocity).  
When not in the strict strong shock regime, the density jump will be 
slightly less than $(\gamma + 1)/(\gamma - 1)$.  The temperature is 
a function of the pressure (e.g. $P^{1/4}$ for a radiation dominated 
gas).

Figure 4 shows the temperature evolution of two different zones
for all 10 explosion simulations: Fig. 4a,4b show the results of the
23\,M$_\odot$ binary, 40\,M$_\odot$ progenitors respectively.  The
rightmost panel shows the temperature evolution of two mass zones for
our strong explosion.  When the shock hits the matter, its temperature
increases abruptly as we expect from our shock jump conditions.
Material further out reaches lower peak temperatures.  The leftmost
panel shows the same mass zones in our weak explosion.  Because the
explosion is weaker, the peak temperature is much lower.  The middle 3
panels show the evolution of our weak plus strong explosions with the
total delay decreasing as we move from left to right.  If the delay is long,
the matter expands so much after the first weak explosion that the
second explosion is unable to heat the material significantly.  As the
delay shortens, the peak temperature approaches that of the single
strong shock.  This peak temperature plays a crucial role in
determining the nucleosynthetic yield.  Recall that the total energy
in all the weak+strong and the strong explosions are nearly
identical.  So although the energetics of the explosions will all
appear identical, their yields will be quite different.

Figure 5 shows the corresponding density evolution of all our
simulations.  Note the near constant density jump for the weak and
strong explosions.  The slight difference in density jump occurs
because we are approaching, not at, the strong shock limit.
In general for the weak+strong shock models, the material expands
considerably after the first shock and its density after the strong
shock hits is increased, but not higher than its peak value. The
exceptions are the short delays, for which the peak density can be
higher than what would be obtained from a single strong explosion.

\section{Nucleosynthetic Yields}

The nucleosynthetic yield of the ejecta is determined by the initial
abundance and the temperature, density, and electron (or neutron)
fraction evolution. The small nuclear network used in the simulation
itself assumes a fixed neutron excess.  We then post-process the
ejecta in our hydrodynamical explosions using a 524 element nuclear
network.  The initial abundance is set by the stellar models (again,
see Young et al. 2006), which also sets the electron fraction we use
for the hydro simulations. For our models, which leave behind large
remnants, the initial electron fraction is between 0.497 and 0.5.

Figure 6 shows the nickel, titanium, silicon, oxygen, and carbon mass
fraction profiles of our ejecta as a function of enclosed mass.
Actually predicting the yields of our complex temperature and density
evolutionary tracks is very difficult, but we can predict some trends
based on the fact that the peak temperature for a piece of matter
decreases for longer delays in the explosion, first and foremost being
that nuclear burning should be less and we should expect a lower
production rate of the heavier explosive elements like $^{56}$Ni and
$^{44}$Ti.  This trend is compounded by the fact that, for longer
delays, more material falls back and less inner material (material
that typically forms these heavy elements) is ejected.  And indeed, 
this is the trend we see in our simulations - longer delays in the 
explosion produce less $^{56}$Ni and $^{44}$Ti.

The detailed yields of all our elements heavier than hydrogen are
listed in tables 2 and 3.  Beyond the fact that the weak plus strong
explosions with long delays produce much less $^{56}$Ni and $^{44}$Ti,
we highlight just a few other features of these models.  First off,
although there are mass elements in the star that produce high
$^{44}$Ti to $^{56}$Ni, the composite yields lead to ratios of
$^{44}$Ti to $^{56}$Ni that are generally lower than the solar
abundance ratio of the stable daughter products of these elements:
$^{44}$Ca to $^{56}$Fe.  

As with Maeda \& Nomoto (2003), we find the production of $A\gtrsim60$
elements (e.g. $^{64}$Zn and $^{59}$Co) is high relative to iron.  In
the case of our 23\,M$_\odot$ models, the ratio of $^{64}$Zn/$^{56}$Fe
can be 10 times solar.  In general, the ratio decreases with longer
delays.  The enhanced production of these elements is not so extreme
in our 40\,M$_\odot$ models, and the difference between stellar progenitor 
are much greater than the differences caused by our different explosion 
scenarios.  Such abundances are hence better suited to differentiated 
progenitors, and not the delay in the explosion model.

We also specifically studied the production sites of calcium and
titanium.  Figure 7 shows the abundance fraction of stable calcium and
titanium as a function of enclosed mass for 4 models: 23strong,
23WS1.0, 40Strong, 40WS4.8.  On average, the iron is most centrally 
concentrated with the mean of the titanium further out, and the 
mean of the calcium production even further out.  But these regions 
all overlap.  For the long delays, it appears that the distributions 
of all the elements are more centrally located.

A few other points are worthy of note in the nucleosynthesis. Material
initially at roughly the neutron excess of the solar Fe peak can end
up in the ejecta, especially if we include explosions from rapid
accretion onto neutron stars.  In addition, delays of seconds after a
weak explosion can result in matter remaining at high density long
enough for weak interactions to operate and increase the neutron
excess before the strong shock passes through. In the strong
explosions and short delay weak+strong explosions the peak
temperatures are very high.  In most regions, the material is
dissociated into $\alpha$ particles and then burned back up into the
heavy element abundances we observe. In the innermost regions of these
explosions the radiation entropy $S_R > 1.0\times 10^8$, and the
material is dissociated into proton and neutrons before burning
back. This material is a small fraction of the star and so plays
little role in the observed yields but exhibits interesting nuclear
physics. Note also that nearly all of the helium in all of our models
(except our single weak explosions) is burned into heavier elements
(mostly carbon and oxygen).  Even though these stars collapsed as
helium stars, the strong explosions would make them appear to be
helium poor stars.

Lastly, we show the velocity histogram of the different elements 
for a few of our explosions.  The nickel and titanium velocity 
profiles are much more narrow than the silicon, carbon, and oxygen 
profiles. However, the velocities are all high (as we would expect from 
our 13-14$\times 10^{51}$\,erg explosions).  These velocity profiles 
are probably not a good representation of hypernova velocity profiles, 
which are strongly affected by the asymmetries in the explosions.

\section{Conclusion}

A collapsing star can form black holes in two ways: a direct collapse
with no supernova explosion or the delayed collapse caused by fallback
accretion after a weak supernova explosion.  The collapsar engine for
gamma-ray bursts and hypernovae can work using either of these black
hole formation mechanisms.  We have found in this paper that the
nucleosynthetic yields of these two mechanisms, if dominated by
explosive nucleosynthesis, will be quite different.  These differences
can be used to distinguish the black hole formation scenario, and
ultimately, the progenitors of these outbursts.  We have focused on
the $^{56}$Ni yield.  For direct collapse black holes, or black holes
formed shortly ($<$1\,s) after a weak explosion, the nickel yield
tends to be much higher than the yield produced by black holes formed
after a considerable delay ($>1$\,s).  Hypernovae with high nickel 
yields (e.g. SN 1998bw) argue for short delays or direct collapse.  

There are a number of effects that can flatten out this trend.
Explosions are asymmetric and the asymmetries may lead to a lower
nickel yield in the strong explosions.  Also the disk itself may
contribute significantly to the nickel yield by driving winds
producing more nickel in the explosions with long delays.  The yield
also depends sensitively on the exact progenitor.  But we suspect
these complexities will not completely wash out the basic trend of the
decreasing nickel production with an increase in the delay between
weak and strong explosion.  Especially near the $20$\,M$_\odot$
progenitor limit, we see a range of nickel yields in supernovae:
compare the 0.32\,M$_\odot$ of $^{56}$Ni predicted for SN2005bf
(Tominaga et al. 2005) against the $\sim 4$\,M$_\odot$ of $^{56}$Ni in
SN1999as (Deng et al. 2001).

We also notice that for our particular progenitors, the stars are
sufficiently compact and the hypernova explosion energies are so large
that, even though these stars collapse as helium stars, their helium
is almost completely burned into heavier elements.  So a helium-rich
star (technically believed to be a Type Ib progenitor) can explode as
a helium-poor star (or Type Ic supernova).  Again, asymmetric
explosions and different progenitors can alter this result, but it
suggests that possibly the fact that most hypernovae appear to be type
Ic supernovae may not mean that the progenitor star has lost most of
its helium envelope at collapse.

{\bf Acknowledgments} We would like to thank many useful conservations
with Dave Arnett, Paolo Mazzali, Ken Nomoto, and Kei'ichi Maeda.  This
work was funded in part under the auspices of the U.S.\ Dept.\ of
Energy, and supported by its contract W-7405-ENG-36 to Los Alamos
National Laboratory, by a DOE SciDAC grant DE-FC02-01ER41176, a NASA
grant SWIF03-0047, and by National Science Foundation under Grant
No. PHY99-07949.

{}

\newpage

\begin{deluxetable}{lccccc}
\tablewidth{0pt}
\tablecaption{Explosion Models}
\tablehead{
  \colhead{Model}
& \colhead{t$_{\rm delay}$\tablenotemark{a}}
& \colhead{$\dot{E}$\tablenotemark{b}}
& \colhead{t$_{\rm Heating}$\tablenotemark{c}}
& \colhead{E$_{\rm Explosion}$\tablenotemark{d}}
& \colhead{M$_{\rm Remnant}$\tablenotemark{e}} \\
\colhead{Name}
& \colhead{(s)}
& \colhead{($10^{54}$erg\,s$^{-1}$)}
& \colhead{(ms)}
& \colhead{(10$^{51}$\,erg)}
& \colhead{(M$_\odot$)}
}
\startdata
23Weak & 0 & 1  & 4 & 0.25 & 2.1 \\
23WS0.7 & 0.7\tablenotemark{f} & 8 & 3 & 15 & 1.4 \\
23WS1.0 & 1.0\tablenotemark{f} & 8 & 3 & 15 & 1.6 \\
23WS2.6 & 2.6\tablenotemark{f} & 8 & 5 & 16 & 1.8 \\
23Strong & 0 & 8  & 4 & 14 & 1.4 \\
40Weak & 0 & 2  & 24 & 0.1 & 6.3 \\
40WS1.0 & 1.0\tablenotemark{f} & 15  & 15 & 16 & 2.4 \\
40WS4.8 & 4.8\tablenotemark{f} & 15  & 16 & 16 & 3.0 \\
\enddata

\tablenotetext{a}{Delay between the beginning of simulation and the
time we begin to deposit energy.}  \tablenotetext{b}{Energy deposited
in the inner 15 zones of the simulation.  For the delayed explosions,
we use the heating rate for the weak explosion and then, after a
delay, add the heating rate listed in the table.}
\tablenotetext{c}{Time that we deposit the energy.}
\tablenotetext{d}{Total kinetic energy in the explosion when the shock
hits the edge of the star.}  \tablenotetext{e}{Mass of material that
will not escape the gravitational potential well of the remnant.  Some
of this material is still moving outward at the end of the simulation,
but its velocity is less than the escape velocity.}
\tablenotetext{f}{The delay times for the 23\,M$_\odot$ models
correspond to the onset of accretion (23WS0.7), a 1.6\,M$_\odot$
remnant (23WS1.0), and a 1.7\,M$_\odot$ remnant (23WS2.6).  The delay
times for the 40\,M$_\odot$ models correspond to an early explosion at
1.0s (40WS1.0), the onset of accretion (40WS4.8), and a 2.6\,M$_\odot$
remnant (23WS2.6).}

\end{deluxetable}
\clearpage

\begin{deluxetable}{lccccc}
\tablewidth{0pt}
\tablecaption{Detailed Yields (M$_\odot$) Before Decay of Radioactive Species for 40\,M$_\odot$ Progenitor}
\tablehead{
  \colhead{Species}
& \colhead{Weak}
& \colhead{WS6.8s}
& \colhead{WS4.8s}
& \colhead{WS1.0s}
& \colhead{Strong}}

\startdata

$^{ 3}$He & 2.36E-25 & 1.00E-17 & 4.92E-17 & 3.21E-15 & 1.56E-15 \\    
$^{ 4}$He & 4.59E-03 & 3.16E-02 & 7.27E-02 & 1.28E-02 & 8.97E-02 \\    
$^{ 6}$Li & 3.15E-21 & 1.78E-19 & 1.69E-19 & 1.91E-19 & 1.88E-19 \\    
$^{ 7}$Li & 1.51E-25 & 1.79E-21 & 4.57E-21 & 1.23E-20 & 7.01E-21 \\    
$^{ 7}$Be & 6.76E-24 & 9.14E-21 & 8.90E-21 & 8.45E-21 & 6.26E-21 \\    
$^{ 8}$Be & 1.93E-25 & 1.99E-06 & 1.35E-06 & 1.20E-06 & 2.41E-06 \\    
$^{ 9}$Be & 1.66E-23 & 1.60E-21 & 7.94E-22 & 6.28E-22 & 2.25E-22 \\    
$^{ 8}$B  & 1.52E-41 & 1.17E-33 & 7.96E-34 & 1.81E-36 & 2.08E-38 \\
\enddata

\end{deluxetable}
\clearpage

\begin{deluxetable}{lccccc}
\tablewidth{0pt}
\tablecaption{Detailed Yields (M$_\odot$) Before Decay of Radioactive Species for 23\,M$_\odot$ Progenitor}
\tablehead{
  \colhead{Species}
& \colhead{Weak}
& \colhead{WS6.8s}
& \colhead{WS4.8s}
& \colhead{WS1.0s}
& \colhead{Strong}}

\startdata

$^{ 3}$He & 2.74E-15 & 1.10E-13 & 1.11E-13 & 9.08E-14 & 1.11E-14 \\    
$^{ 4}$He & 4.88E-04 & 1.23E-03 & 1.12E-01 & 1.82E-01 & 1.55E-01 \\    
$^{ 6}$Li & 1.92E-21 & 5.62E-25 & 1.18E-25 & 5.87E-26 & 8.61E-21 \\    
$^{ 7}$Li & 2.58E-25 & 1.41E-24 & 1.58E-24 & 1.55E-24 & 3.32E-21 \\    
$^{ 7}$Be & 5.45E-21 & 1.96E-18 & 9.47E-18 & 7.49E-18 & 8.79E-19 \\    
$^{ 8}$Be & 2.92E-11 & 4.56E-06 & 7.21E-06 & 4.85E-06 & 4.71E-06 \\    
$^{ 9}$Be & 1.67E-25 & 2.57E-28 & 1.55E-27 & 7.93E-27 & 1.65E-21 \\    
$^{ 8}$B  & 1.37E-33 & 3.56E-27 & 3.30E-24 & 4.32E-24 & 4.11E-25 \\
\enddata

\end{deluxetable}
\clearpage

\begin{figure}
\plotone{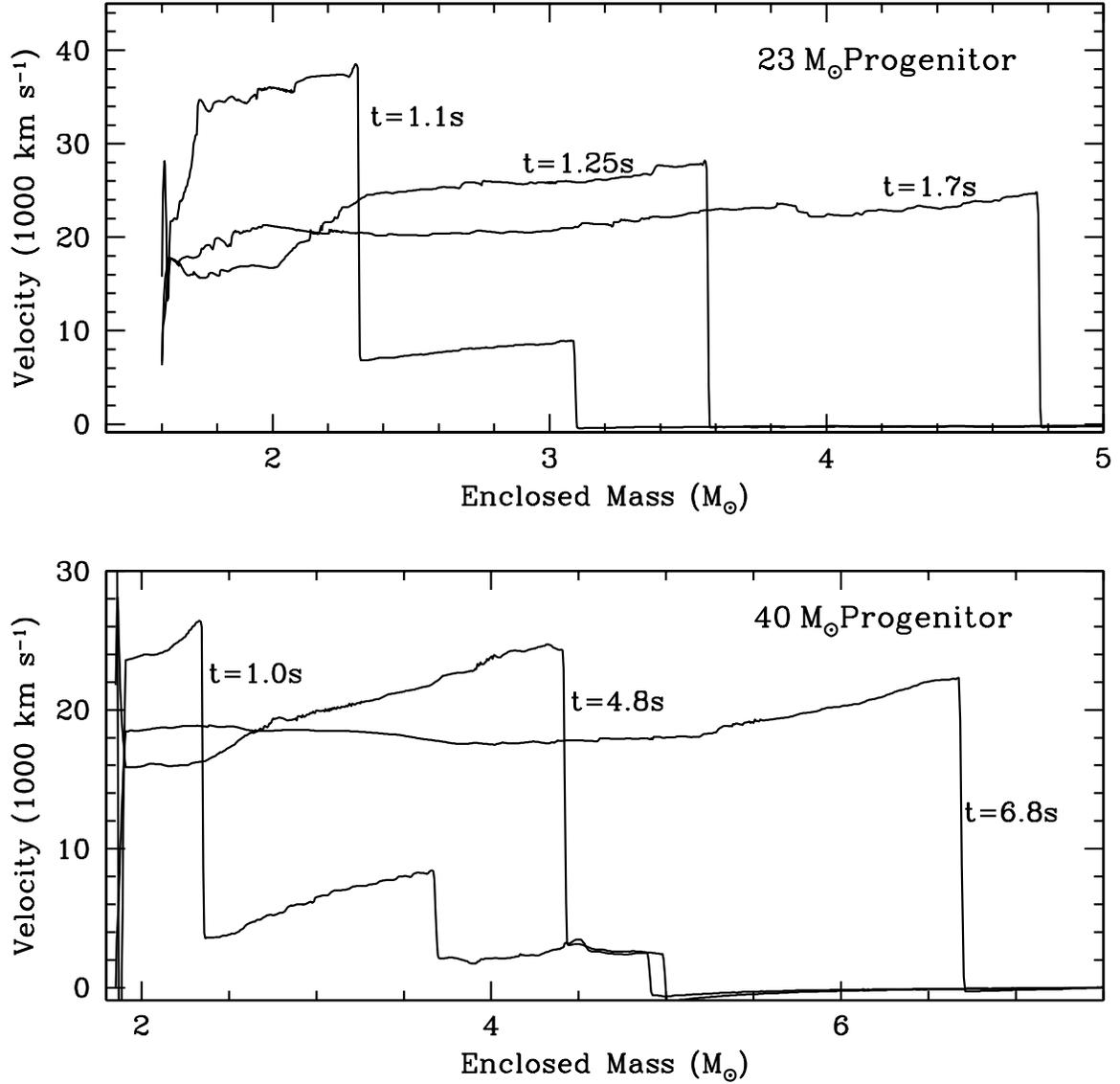}
\caption{Velocity versus enclosed mass for the 23\,M$_{\odot}$ (top) and
40\,M$_{\odot}$ (bottom) progenitor with strong explosion energies
(14,15 $\times 10^{51}$\, erg for the 23\,M$_{\odot}$,40\,M$_{\odot}$
progenitors).  Note that the peak velocities as the shock reaches 
the edge of the star still exceed 20,000\,km\,s$^{-1}$.}
\label{fig:vvsmstrong}
\end{figure}
\clearpage

\begin{figure}
\plotone{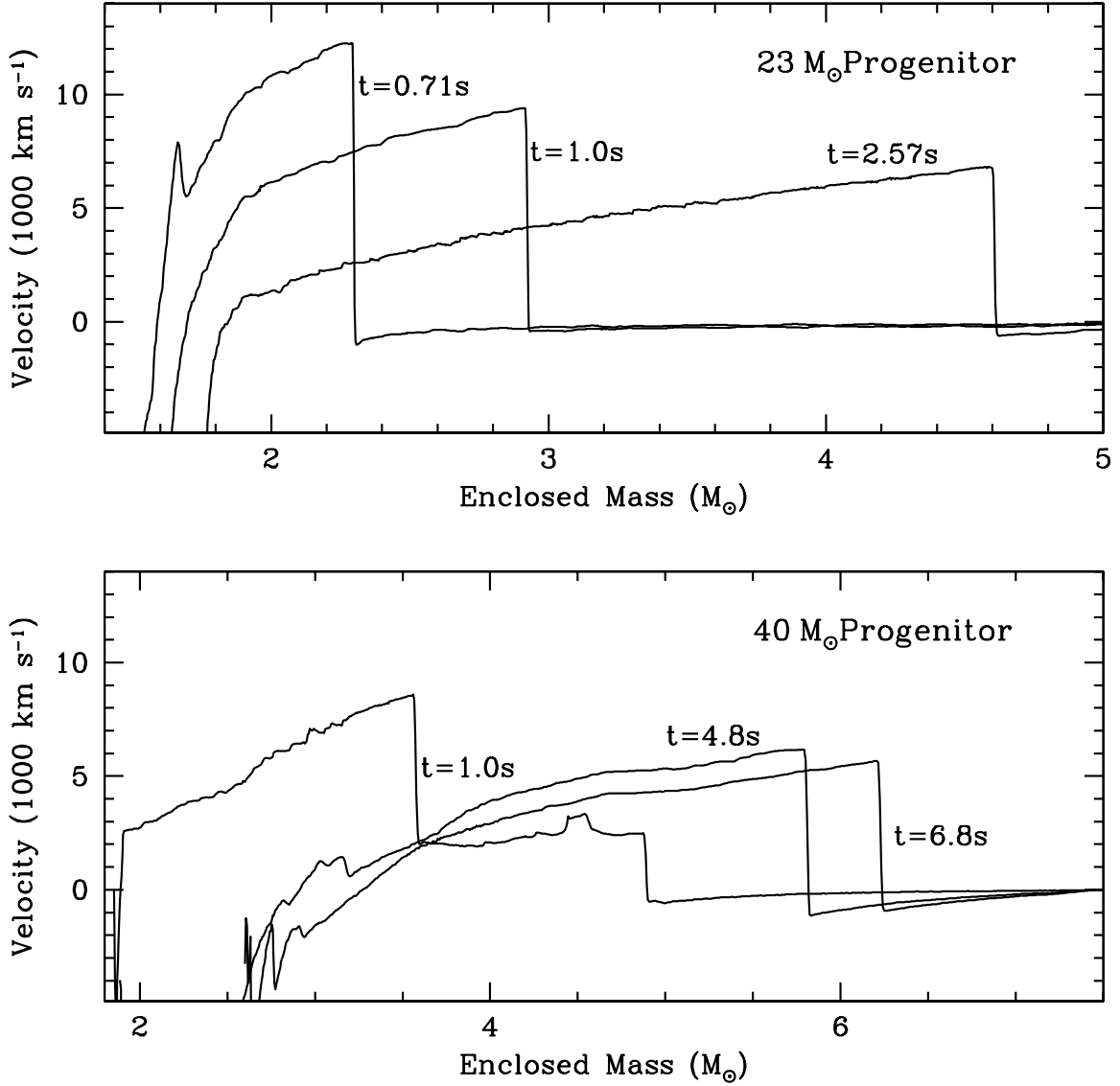}
\caption{Velocity versus enclosed mass for the 23\,M$_{\odot}$ (top) and
40\,M$_{\odot}$ (bottom) progenitor with weak explosion energies
(0.25,0.1 $\times 10^{51}$\, erg for the 23\,M$_{\odot}$,40\,M$_{\odot}$
progenitors).  These correspond to the 23weak and 40weak simulations.
We show the velocities for 3 snapshot times, corresponding to the time
at which we take these models and introduce a new explosion: from left
to right, we have the initial conditions for models 23WS0.7, 23WS1.0,
23WS2.6 (top panel) and models 40WS1.0, 40WS4.8, 40WS6.8 (bottom
panel).}
\label{fig:vvsmweak}
\end{figure}
\clearpage

\begin{figure}
\plotone{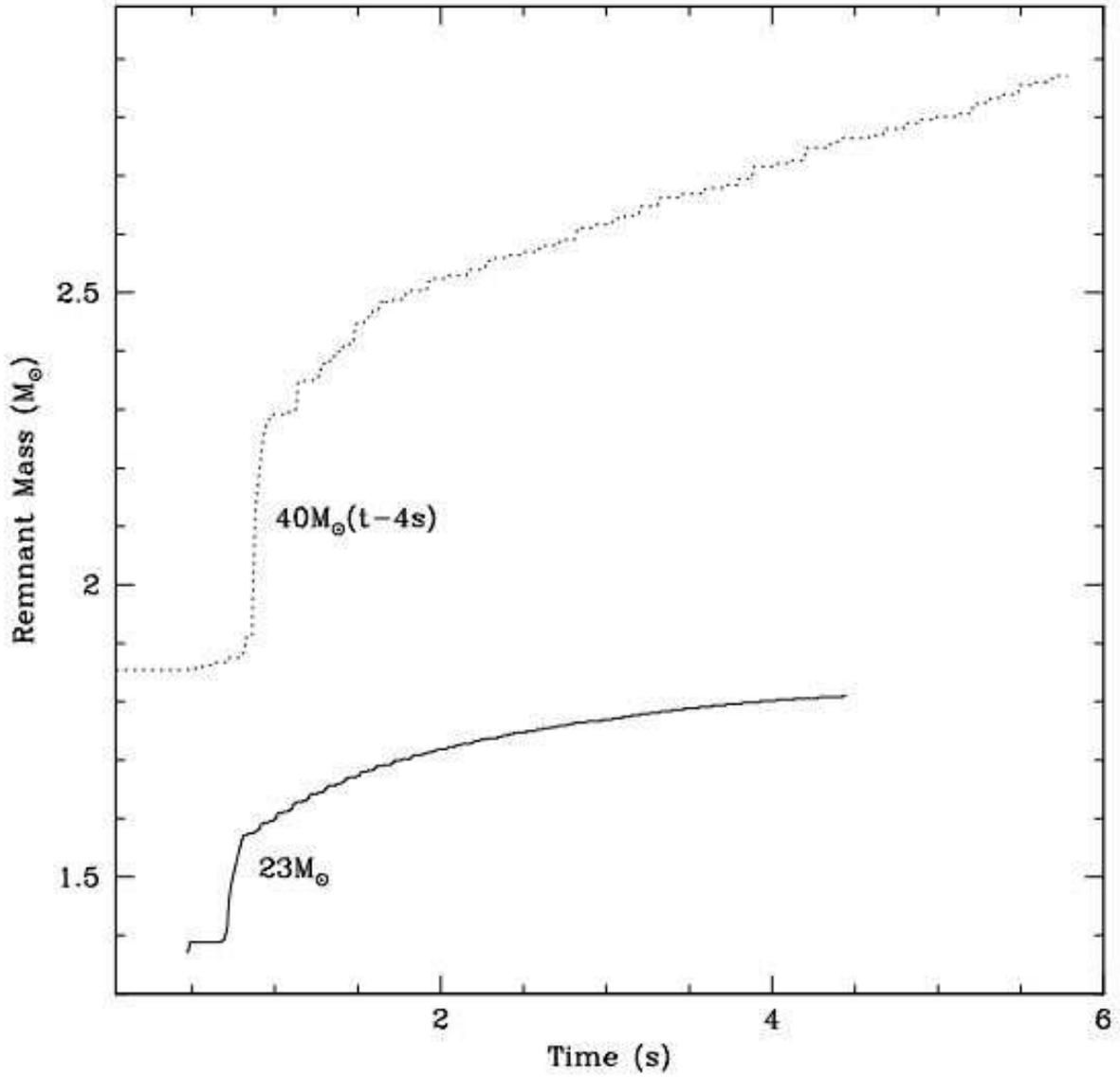}
\caption{Mass of the compact remnant as a function of time for weak
explosions of the 23\,M$_\odot$ (solid) and 40\,M$_\odot$ (dotted)
stars.  This is the mass of matter actually added to the compact
remnant.  The total amount of mass bound to the remnant (that will
ultimately accrete onto the remnant) is much larger.}
\label{fig:mass}
\end{figure}
\clearpage

\begin{figure}
\plottwo{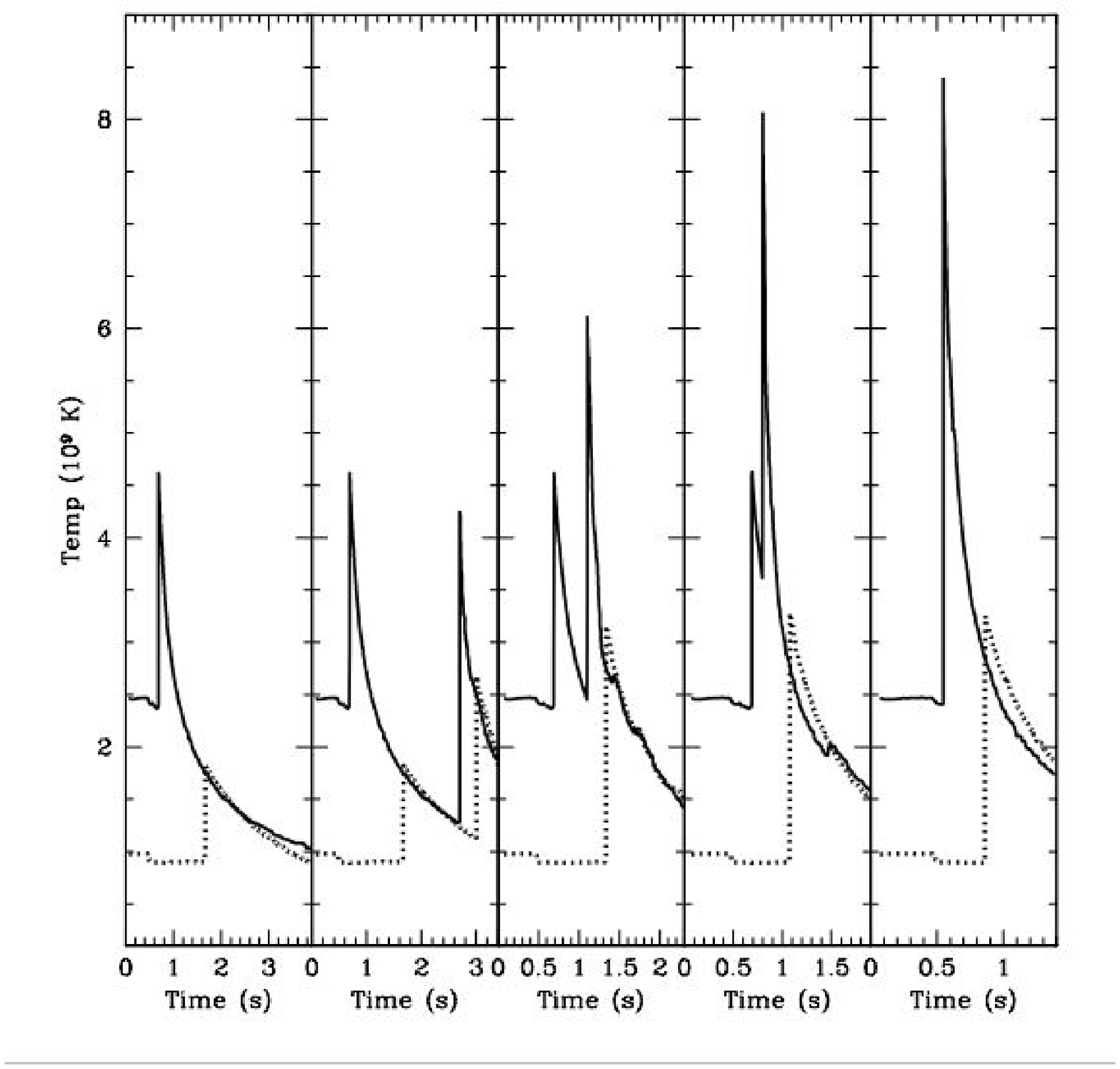}{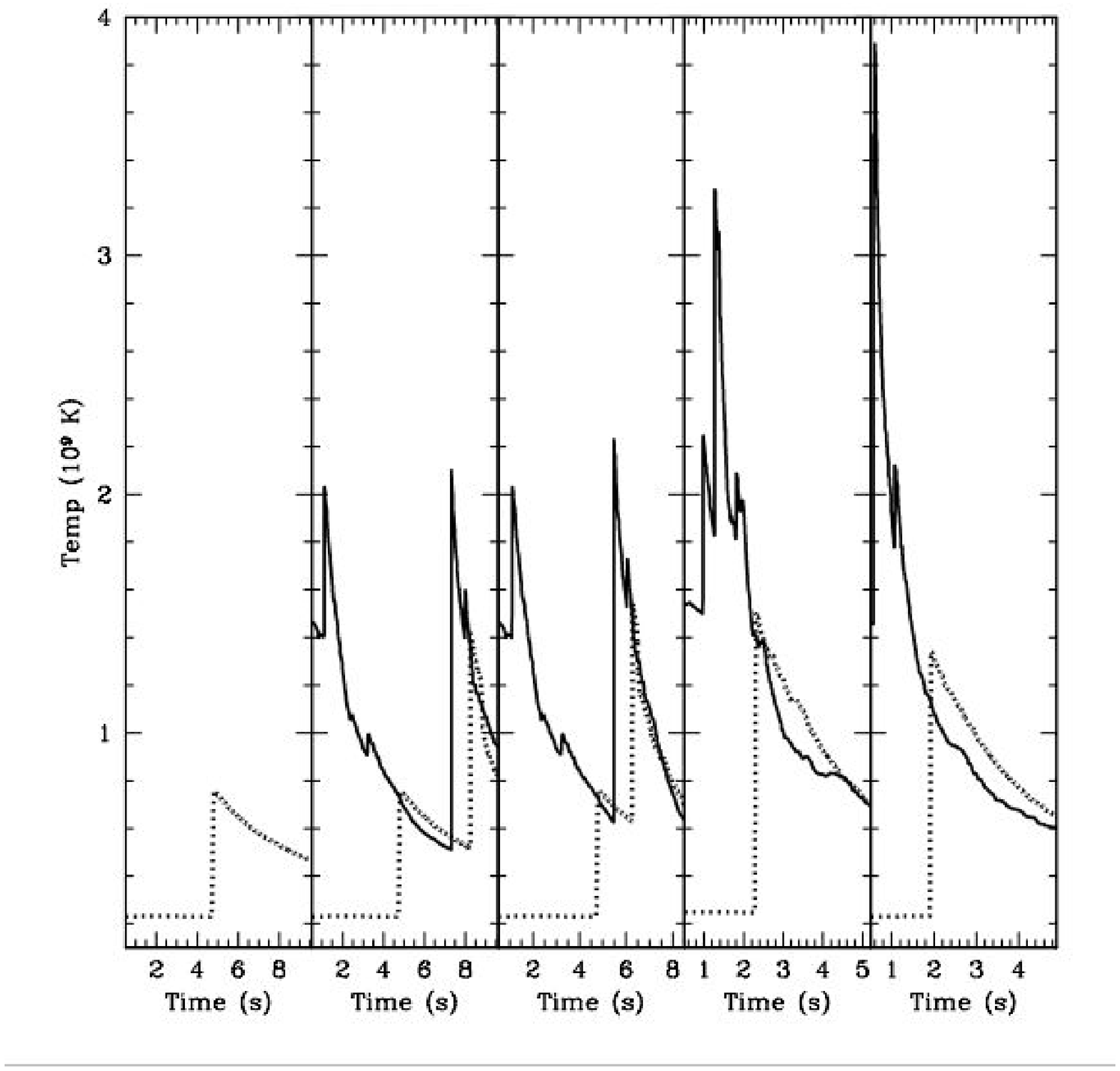}
\caption{Temperature versus time for 2 mass zones for all our models.
Figure 4a corresponds to the 23\,M$_{\odot}$ models: from left to right,
model 23weak, 23WS0.7, 23WS1.0, 23WS2.6, 23strong.  Figure 4b
corresponds to the 40\,M$_{\odot}$ models: from left to right, model
40weak, 40WS1.0, 40WS4.8, 40WS6.8, 40strong.  Note that the highest 
temperatures are achieved with a single shock.  The weak shock moves 
the material outward, leading to lower peak temperatures when the 
strong shock ultimately strikes the material.  Hence we expect considerably 
less nuclear burning in the models with long delays.}
\label{fig:tempe}
\end{figure}
\clearpage

\begin{figure}
\plottwo{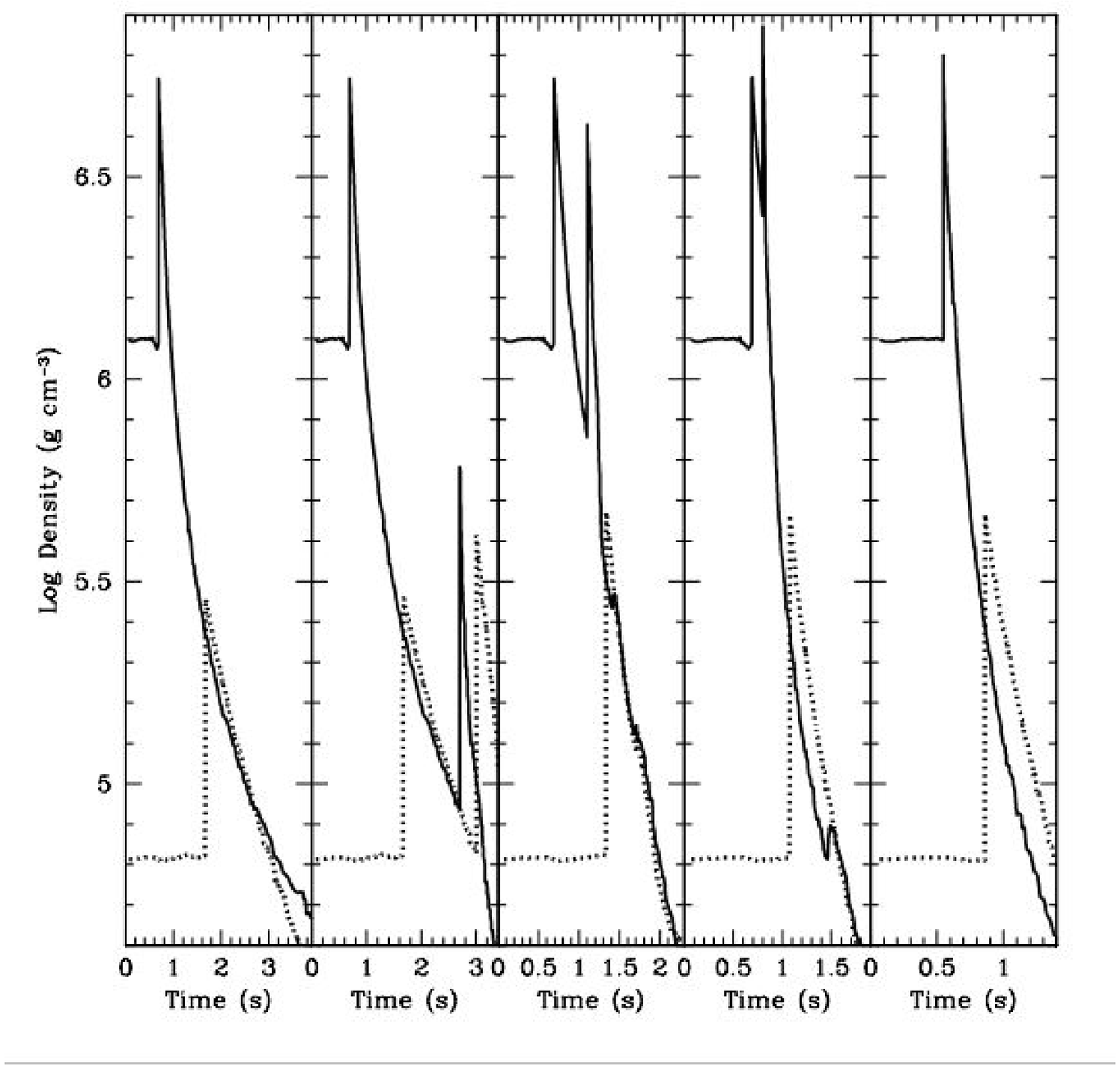}{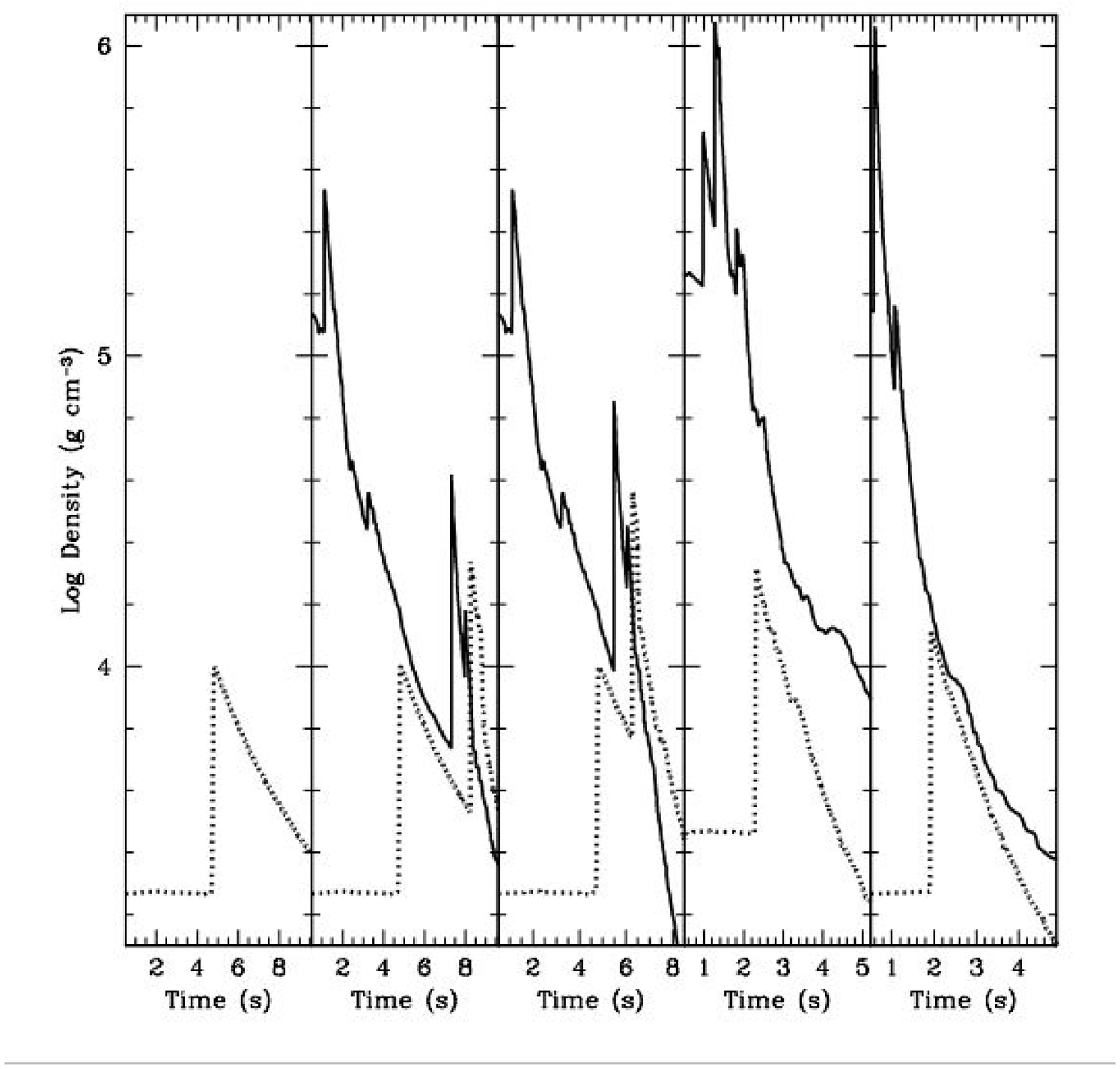}
\caption{Density versus time for 2 mass zones for all our models.
Figure 4a corresponds to the 23\,M$_{\odot}$ models: from left to right,
model 23weak, 23WS0.7, 23WS1.0, 23WS2.6, 23strong.  Figure 4b
corresponds to the 40\,M$_{\odot}$ models: from left to right, model
40weak, 40WS1.0, 40WS4.8, 40WS6.8, 40strong.  Note that because the
density jump is insensitive to the density or strength of the shock
(as long as it is a strong shock), the density of double shocked
material can actually be higher in the case of a short delay.  By
altering both the temperature and density evolution with the delayed
explosions, we are altering the yields for these models.}
\label{fig:rhoe}
\end{figure}
\clearpage

\begin{figure}
\plottwo{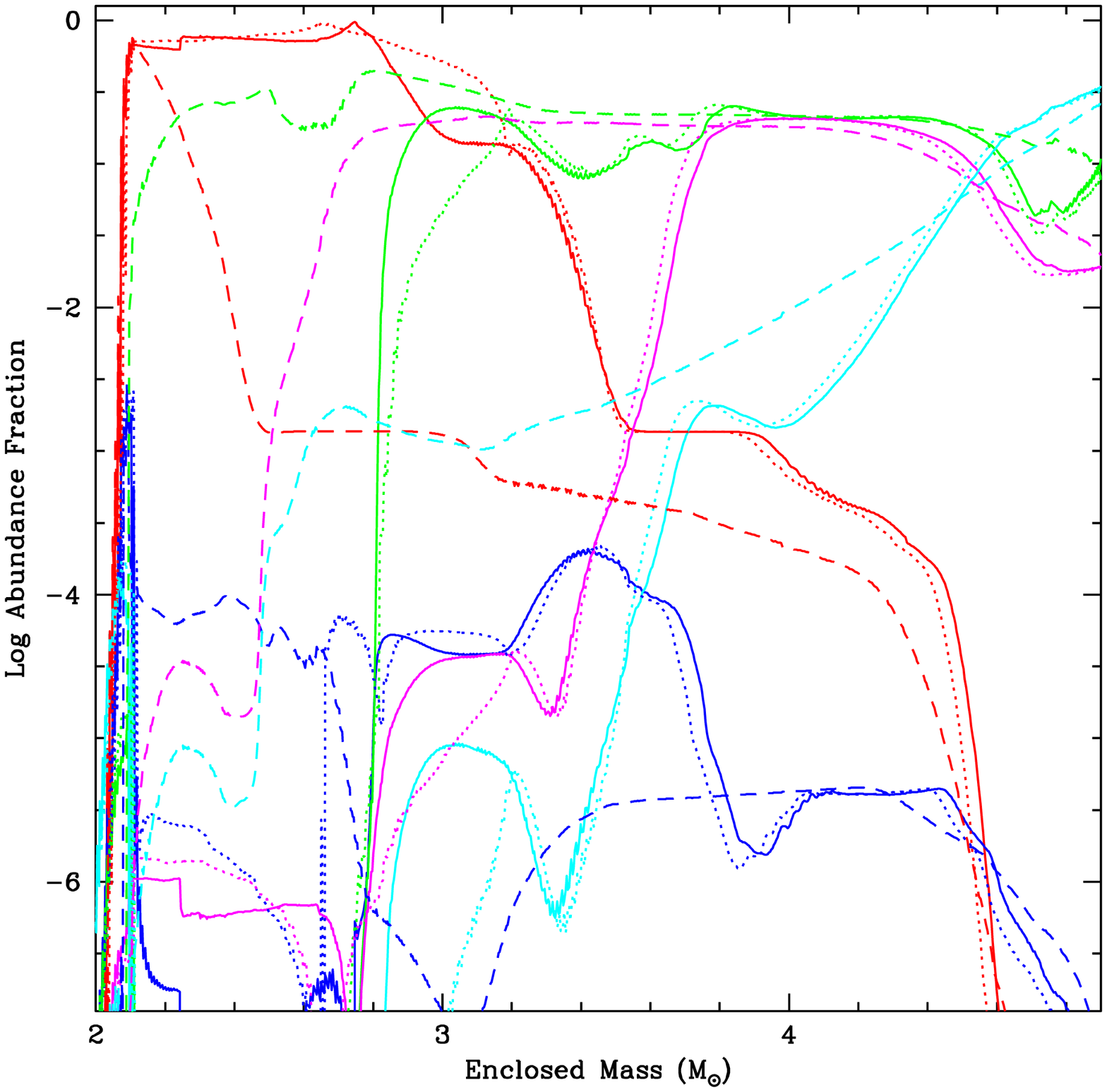}{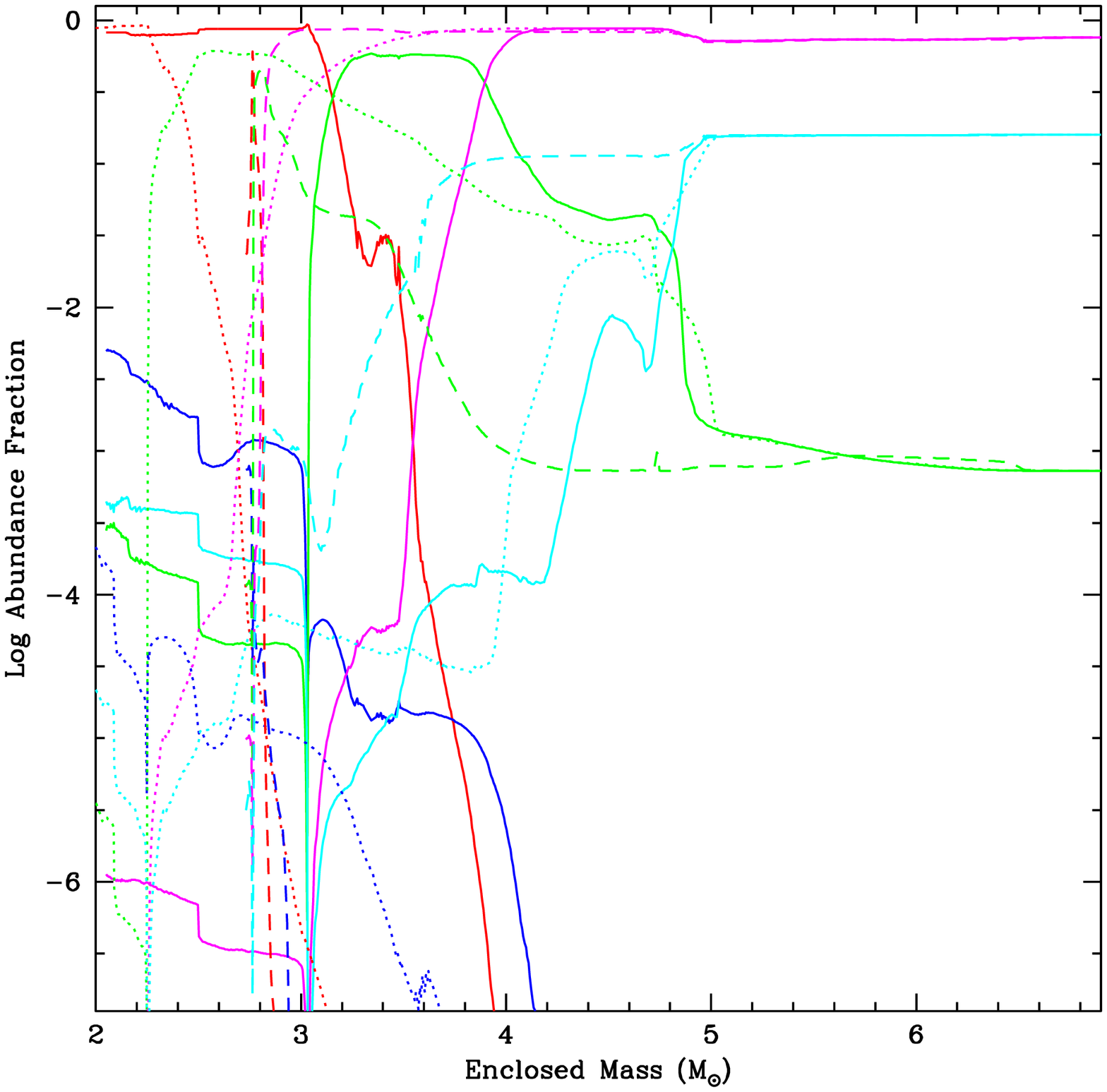}
\caption{Fractional abundance versus enclosed mass 
for 6 of our models:  panel (a) - 23strong (solid), 
23WS1.0 (dotted), 23WS2.6 (dashed) and panel (b) - 
40strong (solid), 40WS1.0 (dotted), 40WS4.8 (dashed).  
The colors denote different abundances prior to decay: 
$^{56}$Ni (red), $^{44}$Ti (blue), $^{28}$Si (green), 
$^{16}$O (magenta), $^{12}$C (cyan).  Note that there 
are zones where the $^{44}$Ti yield exceeds that of the 
$^{56}$Ni.  Also note that nearly all of the He in 
these stars is burned into heavier elements.}
\label{fig:color}
\end{figure}
\clearpage

\begin{figure}
\plotone{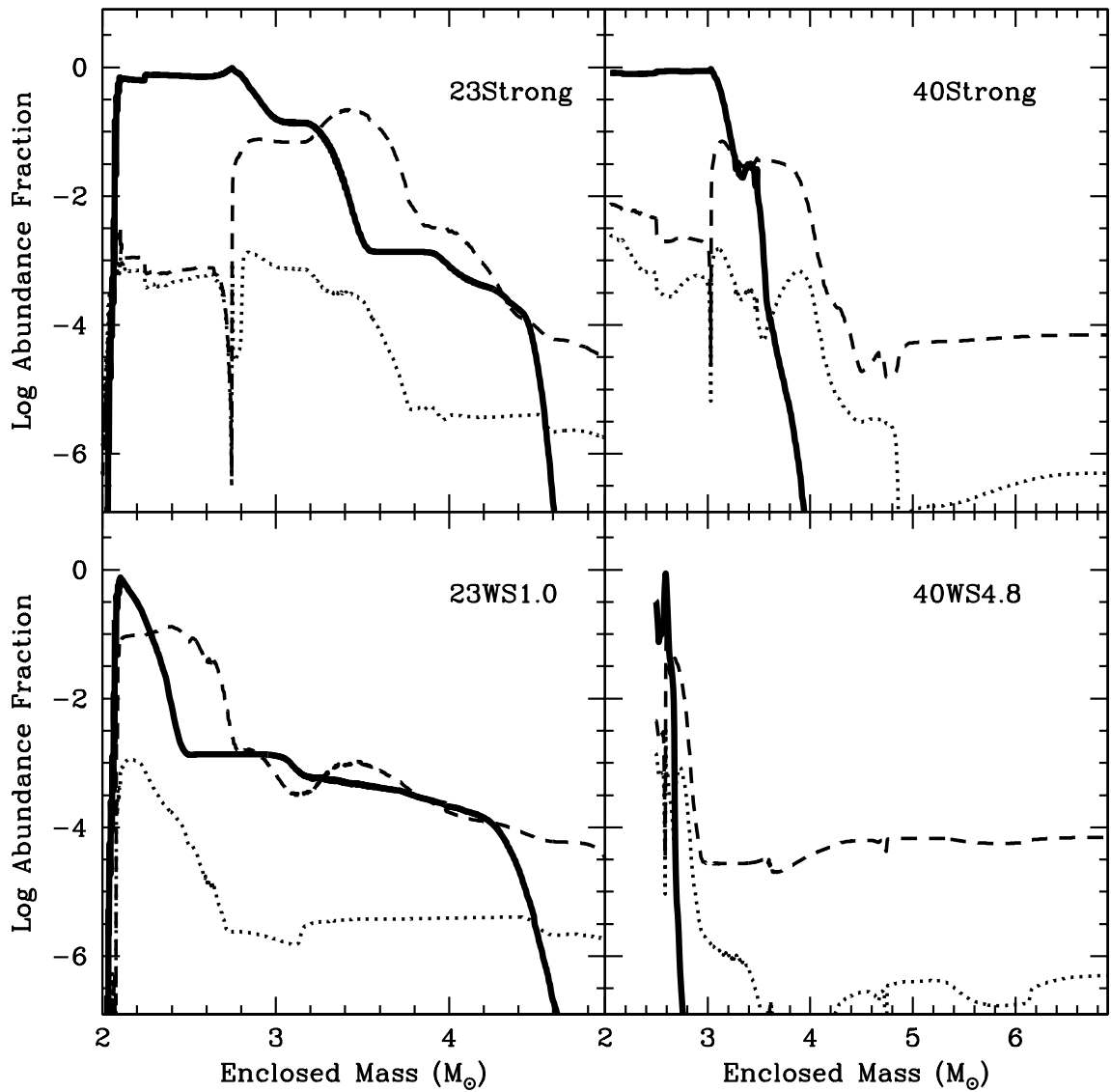}
\caption{Abundance fractions of the stable calcium (dashed), titanium
(dotted) and $^{56}$Fe (solid) versus enclosed mass for 4 of our explosion
models: 23Strong, 23WS1.0, 40Strong, 40WS4.8.  The production site of
the iron is most centrally condensed, followed by titanium and then
calcium.  For strong explosions without delay, a large fraction of the
star is converted into heavy elements.  With long delays, the ejecta
of these heavy elements is confined to the lowest layer of ejected
material.}
\label{fig:cati}
\end{figure}
\clearpage

\begin{figure}
\plottwo{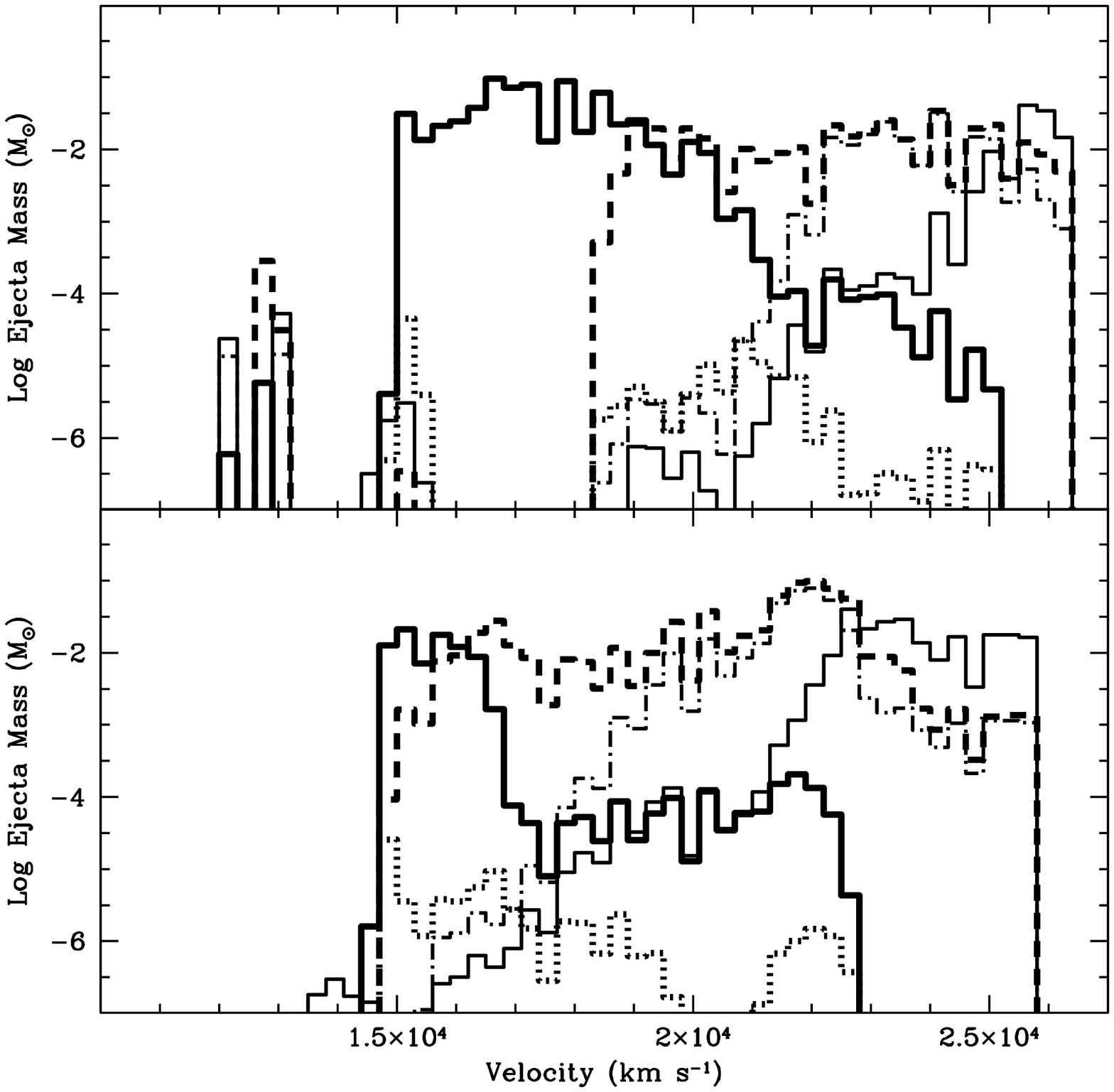}{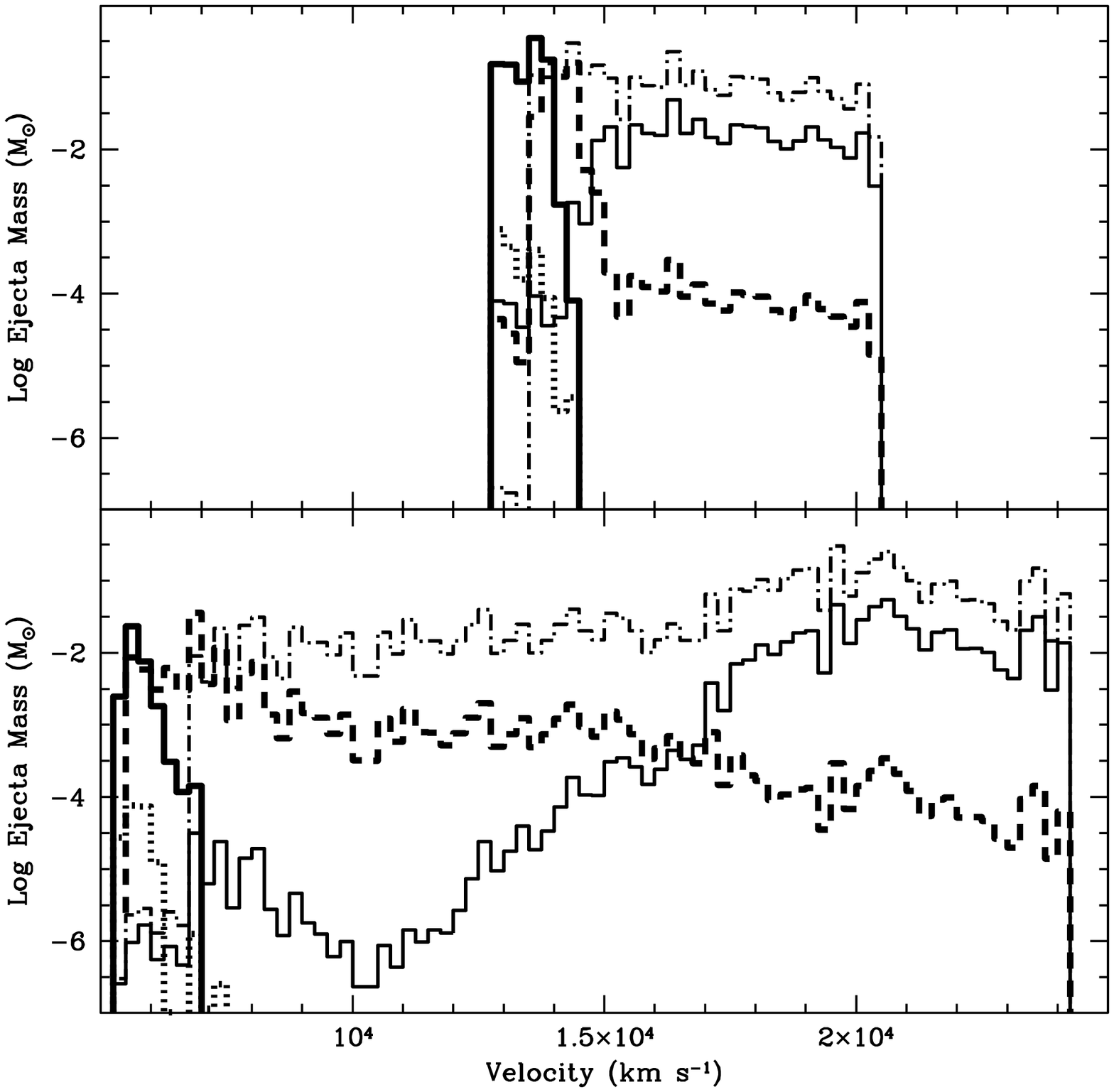}
\caption{Histogram of the mass distribution (in solar masses per bin)
in velocity space for $^{56}$Ni (thick solid), $^{44}$Ti (dotted),
$^{28}$Si (dashed), $^{16}$O (dot-dashed), and $^{12}$C (thin solid).  
The left panels show the results from the 23\,M$_\odot$ models (top -
23strong, bottom 23WS1.0).  The right panels show the results from the
40\,M$_\odot$ models (top - 40strong, bottom 40WS1.0).  No obvious 
trends appear and this distribution is far more dependent upon the 
supernova asymmetry than are the nucleosynthetic yields.
}
\label{fig:veldist}
\end{figure}
\clearpage

\end{document}